\documentclass{valhalla}
\usepackage{amsmath,amssymb,array,longtable,calc,pdflscape,textcomp}
\usepackage{xurl}
\setcitestyle{super,open={},close={}}
\hypersetup{pdftitle={ADMET-EvO: sustained self-evolution across heterogeneous ADMET prediction tasks}}
\DeclareFontShape{T1}{valhallasans}{b}{n}{<->ssub*valhallasans/m/n}{}
\DeclareFontShape{T1}{valhallasans}{bx}{n}{<->ssub*valhallasans/m/n}{}
\DeclareFontShape{T1}{valhallasans}{m}{it}{<->ssub*valhallasans/m/n}{}

\author[1, 2, *]{Yiling Zhou}
\author[1, 3, *]{Yilin Wang}
\author[1, 2]{Jianmin Wang}
\author[1, 2]{Heqin Zhu}
\author[1]{Zirui Wang}
\author[3]{Chang-yu Hiesh}
\author[1, 4]{Kejun Ying}
\author[1, \dagger]{Jiaqi Wang}
\author[1, 2, \dagger]{Yuzhi Xu}
\author[3, \dagger]{Tingjun Hou}
\author[1, 2, \dagger]{Odin Zhang}

\affiliation[1]{Valhalla Technology}
\affiliation[2]{The Chinese University of Hong Kong}
\affiliation[3]{Zhejiang University}
\affiliation[4]{Stanford University}

\contribution[*]{Equal Contribution}
\contribution[\dagger]{Corresponding Authors}
\paperlogo{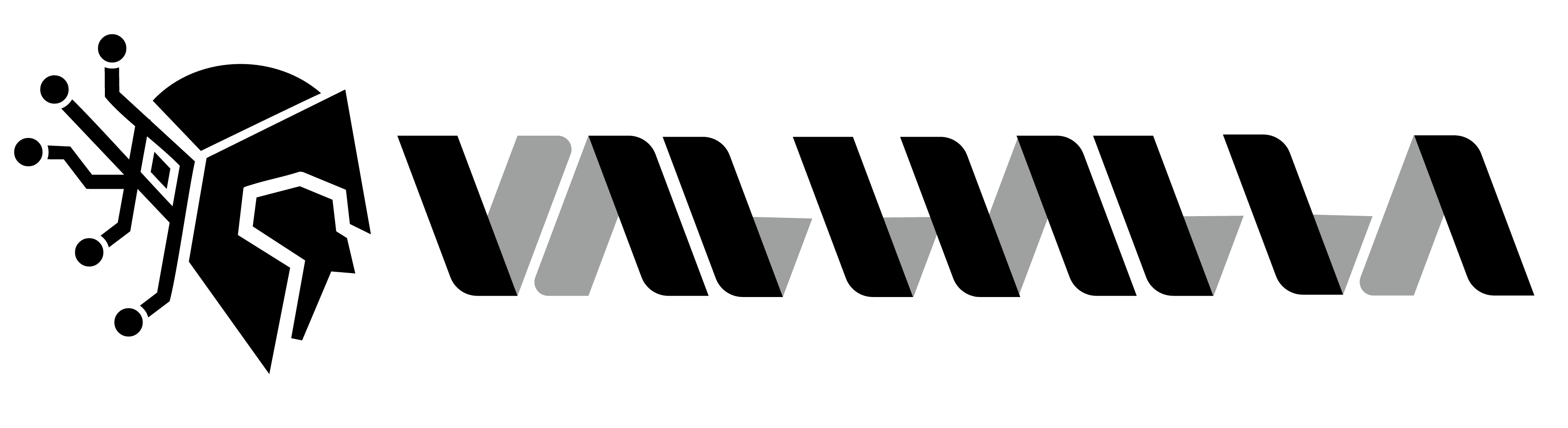}
\papertagline{}
\title{ADMET-EvO: a self-evolving scientific agent for sustained research across heterogeneous tasks}
\abstract{Scientific agents can move beyond automated model building by using accumulated evidence to revise both their questions and experimental strategies. The challenge is sustaining this adaptation across heterogeneous tasks without overfitting decisions to internal validation. Absorption, distribution, metabolism, excretion and toxicity (ADMET) prediction provides a demanding setting across diverse assays, datasets and chemical domains. We therefore developed ADMET-EvO, an evidence-gated agent that formalizes endpoints, generates falsifiable hypotheses and tests interventions across data, feature and model axes. It carries supported, rejected and inconclusive outcomes forward to guide each new cycle. Across the 22-task Therapeutics Data Commons (TDC) ADMET benchmark, ADMET-EvO achieved the highest task-normalized score of 96.77. Evidence-guided selection reduced cumulative fitting time by 72.2\% within a predefined non-inferiority margin. It also formalized 43 toxicity-related tasks and constructed endpoint-specific predictors. Together, these results show how ADMET-EvO can accumulate evidence, revise its strategy and expand its research scope over time.}

\correspondence{odin@valtech.ai; tingjunhou@zju.edu.cn; yuzhixu@valtech.ai; jiaqi@valtech.ai}
\checkdata[Code]{\url{https://github.com/OTeam-AI4S/ADMET-EvO}}
\begin{document}
\maketitle

\section{Introduction}

Self-evolving scientific research extends automated experimentation from isolated optimization runs to a persistent process in which language-model agents iteratively formulate hypotheses, conduct computational experiments, evaluate the resulting evidence, and revise subsequent research decisions within an autonomous closed loop\citep{ref01,ref02,ref03,ref04,ref05}. Progress is therefore represented not by a single optimization trajectory, but by an evolving research program that connects hypotheses, interventions, outcomes, and revisions across successive cycles\citep{ref01,ref03,ref04,ref06}. As evidence accumulates, such systems can refine task definitions, curate and extend datasets, modify representations and predictive models, and redirect exploration toward unresolved questions or poorly covered regimes\citep{ref01,ref05,ref06,ref07}. In this way, self-evolution operates not only on model parameters or individual experiments, but on the research process itself, allowing previous successes and failures to shape both the next experiment and the range of questions the system is able to address\citep{ref01,ref03,ref06}.

Absorption, distribution, metabolism, excretion and toxicity (ADMET) prediction provides a stringent setting for this form of self-evolving research. Historically, computational ADMET research has progressed along two complementary directions. One develops quantitative structure--activity relationship or machine-learning models for individual endpoints\citep{ref08,ref09,ref10}; the other integrates multiple predictors into broadly accessible platforms, including SwissADME, admetSAR, ADMETlab 3.0 and, more recently, ADMET-AI\citep{ref11,ref12,ref13,ref14}. These platforms have substantially expanded the range and throughput of computational ADMET assessment. Their knowledge state, however, remains largely release-bound: endpoint definitions, training datasets and predictive models are revised periodically by developers rather than continuously re-examined as new evidence becomes available. Consequently, broad endpoint coverage does not by itself provide a mechanism for determining whether a newly identified assay is compatible with an existing task, whether an external dataset expands the relevant chemical space, or whether accumulating evidence warrants revising the task definition or modelling strategy\citep{ref08,ref09,ref10,ref15}. Moreover, constructing a reliable predictor for each endpoint still demands substantial expert effort in task specification, data curation, representation and model selection, hyperparameter optimization, and validation design. Much of this workflow must be repeated whenever the dataset, assay definition, or intended chemical domain changes, limiting scalability and slowing the incorporation of new evidence\citep{ref09,ref10,ref15}.

Recent language-model agents have begun to automate different components of computational drug discovery, but ADMET has generally remained a downstream evaluation tool or a fixed prediction task rather than the subject of a sustained research process\citep{ref16,ref17,ref18,ref19,ref20}. Systems such as LIDDIA and AgentD coordinate existing tools for information retrieval, molecular generation, docking and property prediction\citep{ref16,ref17}. AgentD, for example, retrieves literature and predicts 75 ADMET-related and physicochemical properties using integrated predictors, including the established Deep-PK framework\citep{ref17,ref18}. DrugAgent moves closer to model-level research by iteratively proposing and implementing machine-learning pipelines, but its ADMET evaluation is confined to a single predefined PAMPA classification task and primarily explores data processing, molecular representation and algorithmic choices within that benchmark\citep{ref19}. Closed-loop Auto Research broadens this scope by searching feature, model and external-data interventions across 36 molecular-property endpoints and certifying selected configurations on held-out data\citep{ref20}. Collectively, these studies establish important components of agentic ADMET workflows, yet remain either tool-centric or task-bounded. They do not form a persistent research program that continuously formalizes endpoints, diagnoses task hardness, accumulates both positive and negative evidence, adaptively allocates effort across research actions, and initiates new independently certified cycles as the ADMET evidence landscape evolves.

Here we introduce ADMET-EvO, a self-evolving framework that treats ADMET modelling as a continuing scientific inquiry rather than a collection of isolated prediction tasks. At its core, an autonomous agent acts as the system's scientific brain, acquiring biochemical knowledge about each endpoint and its experimental context before deciding how the problem should be studied. Grounded in this knowledge, ADMET-EvO proceeds through a five-stage research cycle. Endpoint Specification formalizes the task, after which Epistemic Profiling identifies how data scarcity, distribution shift, assay noise and related limitations constrain reliable inference. Hypothesis Branching converts this profile into competing explanations, which Multiaxial Perturbation tests through controlled interventions along the data, model and feature axes. Falsification \& Refinement then evaluates each hypothesis against independent evidence, classifies the outcome as supported, rejected or inconclusive, and updates the next research cycle accordingly. Applied across 22 TDC tasks\citep{ref09}, ADMET-EvO built a cross-endpoint knowledge base and achieved the top task-normalized score of 96.77 and won most head-to-head comparisons with MiniMol\citep{ref21} and Trimole-Hybrid\citep{ref22}, supporting its ability to adapt across heterogeneous predefined endpoints. Ablations across the three intervention axes further showed that the performance gains from agent-selected actions observed on the internal validation set persisted on held-out data, supporting the generalizability of the learned policy. To probe generalization beyond predefined benchmarks, ADMET-EvO autonomously explored previously unspecified ADMET tasks. It identified 43 scientifically meaningful toxicity prediction tasks for iterative modelling and achieved strong predictive performance across this expanded task set. By carrying these outcomes forward, ADMET-EvO learns not only which models work, but which research strategies remain useful across known and newly formulated tasks.

\section{Results and Discussion}

\subsection{ADMET-EvO establishes a self-evolving framework for sustained ADMET research}

\begin{figure}[!htbp]
\centering
\includegraphics[width=\linewidth,height=0.70\textheight,keepaspectratio]{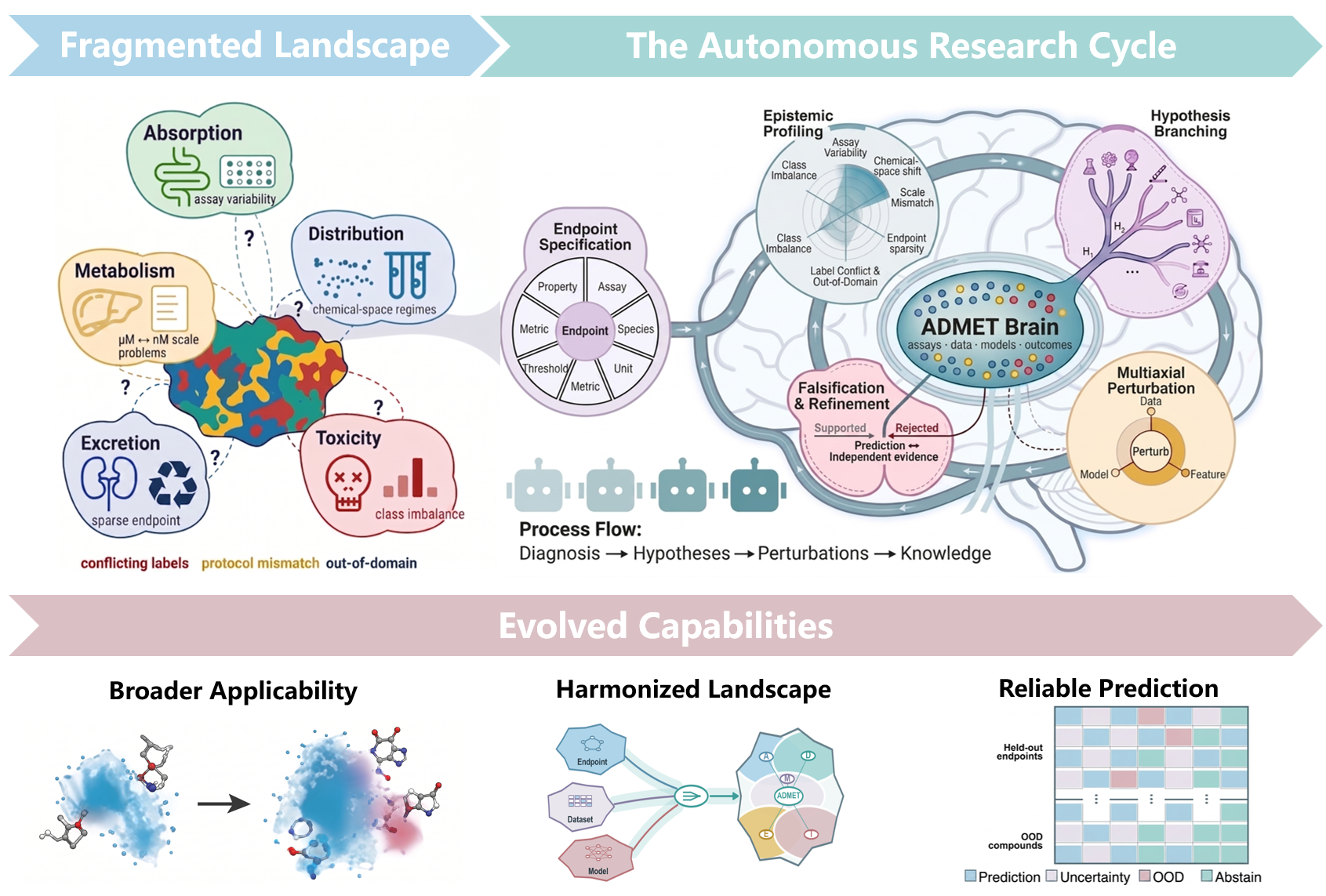}
\caption{\textbf{Overview of the ADMET-EvO framework.} ADMET-EvO transforms fragmented ADMET modelling into a persistent, evidence-driven research process. The ADMET Brain autonomously diagnoses task hardness, generates falsifiable hypotheses, selects targeted perturbations and evaluates outcomes against independent evidence. Supported, rejected and inconclusive findings are retained to guide subsequent cycles, allowing accumulated successes and failures to reshape what is tested next. By evolving the research process itself, ADMET-EvO broadens applicability, harmonizes heterogeneous ADMET tasks and improves prediction reliability.}
\label{fig:1}
\end{figure}

ADMET-EvO organizes autonomous ADMET research into an auditable five-stage cycle comprising Endpoint Specification, Epistemic Profiling, Hypothesis Branching, Multiaxial Perturbation and Falsification \& Refinement (\textbf{Fig. \ref{fig:1}}). Endpoint Specification first establishes the measurement boundary for each endpoint before any modelling decision is made. The resulting specification distinguishes observations that share a broad ADMET label but differ in biological target, semantics, assay system, species, measurement scale, source lineage or split ownership. This boundary prevents nominally related measurements from being combined without evidence of comparability. Epistemic Profiling then assesses how the available evidence constrains modelability and attainable improvement. It jointly examines sample and scaffold support, minority-class counts, conflicting measurements, split overlap, chemical-space shift, local label roughness, validation instability and prior frozen performance. In 22 TDC tasks, the resulting profiles revealed scarcity in DILI, HIA and half-life, whereas class imbalance affected several CYP tasks. Data conflicts or split overlap characterized solubility, clearance and hERG. Pronounced chemical-space shift was also detected in DILI, bioavailability and HIA. These signals were interpreted jointly rather than as independent flags. Low AUPRC or high MAE indicated further headroom for improvement. Hypothesis Branching synthesized these observations into falsifiable explanations for the identified limitations and specified the evidence needed to resolve them.

Guided by these hypotheses, Multiaxial Perturbation converts each prioritized uncertainty into a controlled experiment. Each Research Action specifies a causal hypothesis, falsification test, expected effect, evidence confidence, computational cost and scientific risk. Perturbative actions target one of three axes: data, model or feature. A data-axis intervention modifies only the training evidence or curation policy. A model-axis intervention alters only the estimator, architecture or learning objective. A feature-axis intervention changes only the molecular representation. For each intervention, the other two axes and validation seeds remain fixed, while a matched control enables paired comparisons in the endpoint's native metric. Replication subsequently tests whether an apparent gain persists across multiple validation seeds. The agent may abstain or stop when the expected gain falls below uncertainty or no defensible intervention justifies further computation. External observations remain quarantined until measurement compatibility and source-held-out performance are established. This design isolates intervention effects while maintaining experimental stability and computational efficiency.

Falsification \& Refinement imposes a quantitative evidence gate between exploratory optimization and held-out certification. An intervention is classified as supported only when its mean direction-oriented improvement exceeds 0.005; values below \ensuremath{-}0.005 are rejected, and intermediate effects remain inconclusive. Every verdict is appended to the hash-chained Evidence Ledger and used to update subsequent action priorities. Bayesian shrinkage limits the influence of isolated failures while progressively deprioritizing repeatedly unsuccessful actions. Adaptation is restricted to discovery data, with validation evidence used for selection rather than as the final estimate of generalization. Once a finalist is chosen, its complete recipe is serialized and locked with SHA-256 before the test split is loaded. The frozen recipe is then retrained across at least five seeds and independently assessed by the held-out evaluator. By separating adaptive discovery from held-out certification, ADMET-EvO can continuously revise its research strategy while preserving an independent evidentiary basis for every final claim.

\subsection{ADMET-EvO achieves broadly competitive performance across heterogeneous ADMET endpoints}

The ADMET-EvO framework described above was applied to 22 ADMET endpoints. Final performance was evaluated on the held-out test set (\textbf{Tab. S2}). Among the eight endpoints evaluated by AUROC, five exceeded 90.00\%, led by HIA (100.00\%) and DILI (97.47\%), and values across the full set ranged from 65.59\% to 100.00\%. The remaining five classification endpoints were evaluated using the area under the precision--recall curve (AUPRC) because pronounced class imbalance was identified by the ADMET Brain, making precision--recall analysis more informative than AUROC. Four of these endpoints exceeded 70.00\%, with CYP3A4 Veith reaching 88.68\%. Among the regression endpoints, the strongest rank correlations were obtained for microsomal clearance (Spearman\textquotesingle s \ensuremath{\rho} = 0.69) and VDss (\ensuremath{\rho} = 0.70).

\begin{figure}[!htbp]
\centering
\includegraphics[width=\linewidth,height=0.70\textheight,keepaspectratio]{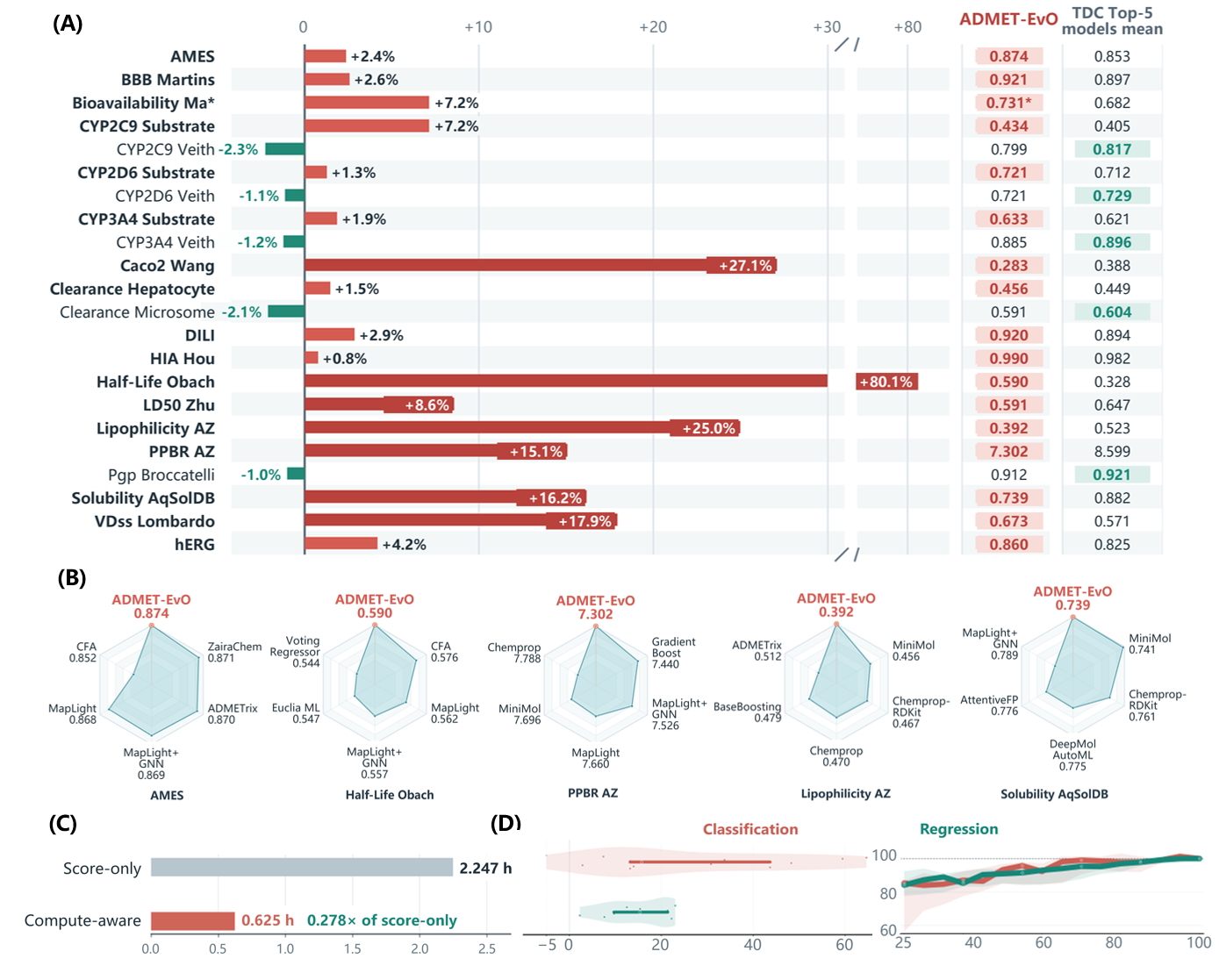}
\caption{\textbf{ADMET-EvO evaluation across 22 ADMET endpoints.} (A) Direction-normalized relative performance against the mean of top-5 TDC models. Positive values favour ADMET-EvO, and scores in their native metrics are shown at right. A broken axis is used for Half-Life. The asterisk identifies the rerun Bioavailability result, used because the value displayed on the TDC website conflicted with multiple published reports. (B) Radar plots for the five endpoints on which ADMET-EvO achieved first-place position in the archived TDC leaderboard. Larger AUROC and Spearman values lie farther from the centre, whereas smaller MAE values lie farther from the centre. (C) Cumulative fitting time of recipes chosen by score-only and compute-aware selection. (D)Association between nearest-training-set similarity and prediction error. Left, distributions of endpoint-level error reduction in the highest-similarity quartile. The horizontal axis gives the percentage reduction from the full-test error, and the vertical axis separates classification (upper) from regression (lower) endpoints. Right, normalized error as coverage expands from the most similar molecules to the complete test set. Errors were normalized within each endpoint. Red denotes classification (n = 13) and green denotes regression (n = 9). Points denote endpoints, diamonds medians and bars interquartile ranges. Shading in the right panel denotes 95\% endpoint-bootstrap confidence intervals.}
\label{fig:2}
\end{figure}

This breadth of performance was accompanied by a recognizable but endpoint-dependent pattern of recipe selection. Tree-based learners appeared in 20 of 22 final recipes (90.91\%), and the Rich2D representation was retained in 18 (81.82\%). Model or representation fusion was selected for 16 endpoints (72.73\%), whereas learned molecular representations were retained selectively for six (27.27\%). One possible explanation for the recurrent Rich2D--tree pairing is that Rich2D combines local substructure, pharmacophore-like and global physicochemical descriptors, providing heterogeneous inputs that tree ensembles may exploit effectively in moderate-data molecular prediction settings. Specialized learned representations were nevertheless retained when supported by endpoint-specific evidence. MapLight+GNN was selected for Bioavailability and CYP2C9 substrate, whereas fine-tuned Uni-Mol2 was used alone for lipophilicity and within fusion models for DILI, Half-Life and hERG. External data were retained for 16 endpoints (72.73\%). Together, the recurrent backbone and selective departures from it show that the iterative search did not simply accumulate model complexity, but adapted the representation, predictor and data source to each endpoint.

The results above emerged from ADMET-EvO\textquotesingle s entire evidence-gated cycle. For direct comparison with established methods, the official TDC task definitions, metrics and data partitions were followed. Selection and weighting used only training and validation evidence, with external data restricted to the training side, before each recipe was frozen, refitted on train plus validation and evaluated on the untouched test set across five seeds. The resulting means were then benchmarked against existing TDC results (\textbf{Fig. \ref{fig:2}A, B}).

\begin{table}[htbp]
\centering
\caption{Task-normalized performance across the 22 TDC ADMET endpoints.}\label{tab:1}
\begin{tabular}{@{}clc@{}}
\toprule\noalign{}
\textbf{Rank}
 & \textbf{Model}
 & \textbf{Task-normalized score}
 \\
\midrule\noalign{}

\textbf{1} & \textbf{ADMET-EvO} & \textbf{96.77} \\
2 & MapLight+GNN\citep{ref23} & 96.64 \\
3 & MapLight\citep{ref23} & 95.05 \\
4 & MiniMol\citep{ref21} & 94.04 \\
5 & MolE\citep{ref24} & 90.06 \\
6 & ContextPred\citep{ref25} & 84.18 \\
7 & AttrMasking\citep{ref25} & 83.29 \\
8 & AttentiveFP\citep{ref26} & 77.96 \\
\bottomrule
\end{tabular}
\end{table}

*Higher scores indicate closer agreement with the best result for each endpoint. All 22 endpoints were weighted equally.

Under this standardized protocol, ADMET-EvO achieved the top task-normalized score of 96.77, followed by MapLight+GNN at 96.64 and MapLight at 95.05 (\textbf{Tab. \ref{tab:1}}). This result shows that ADMET-EvO remained consistently close to the strongest model for each endpoint despite differences in metric scale and optimization direction. ADMET-EvO also outperformed the task-specific mean of the top five TDC models on 17 of 22 tasks (77.3\%; \textbf{Fig. \ref{fig:2}A}). The largest relative gain occurred for Half-Life Obach, where Spearman's \ensuremath{\rho} increased from 0.328 to 0.590 (approximately 80\%). Improvements were also observed for CYP2C9 Substrate (+7.2\% AUPRC), Caco-2 Wang (27.1\% lower MAE), PPBR AZ (15.1\% lower MAE) and VDss Lombardo (+17.9\% in Spearman's \ensuremath{\rho}). Performance on the remaining five tasks was only 1.0--2.3\% below the corresponding top-five mean. Notably, even against MiniMol, whose exact-SMILES filtering carries a potential leakage risk that could inflate its reported performance, ADMET-EvO outperformed it on 13 of 22 endpoints with broader Top-5 coverage (20 versus 18). The same margin held against Trimole-Hybrid, a concurrent, endpoint-specific multimodal model zoo, which ADMET-EvO outperformed on 12 of 22 endpoints with wider Top-5 coverage (20 versus 17). Beyond this broad Top-5 coverage, ADMET-EvO achieved estimated first-place positions on five endpoints in the archived TDC leaderboard snapshot, including AMES (AUROC = 0.874), Half-Life Obach (Spearman\textquotesingle s \ensuremath{\rho} = 0.590), Lipophilicity AstraZeneca (MAE = 0.392), PPBR AZ (MAE = 7.302) and Solubility AqSolDB (MAE = 0.739) (\textbf{Fig. \ref{fig:2}B}). These wins spanned classification, rank-based regression and MAE-based regression, showing that broad competitiveness was accompanied by leading performance across distinct prediction settings.

\subsection{Evidence-guided iteration supports efficiency and interpretable prediction reliability}

Accumulated evidence allowed ADMET-EvO to prioritize lower-cost recipes without materially compromising aggregate performance. Performance was preserved across nearly the entire benchmark, with 20 of 22 endpoints meeting the prespecified non-inferiority criterion and only minor reductions observed for BBB Martins and CYP2D6 substrate. At this performance level, cumulative fitting time for the selected recipes fell from 2.25 h to 0.62 h, a 72.2\% reduction (\textbf{Fig. \ref{fig:2}C}). The largest saving was observed for LD50, where fitting time decreased by 97.3\% while performance remained within the predefined margin. Bioavailability, Solubility and Lipophilicity required 50.4\%, 45.1\% and 12.9\% less fitting time, respectively, while also showing small performance improvements. Thus, accumulated evidence enabled substantially cheaper recipe selection while preserving aggregate predictive performance.

Beyond reducing recipe-selection costs, ADMET-EvO also provided an interpretable signal of where its predictions were most reliable. Prediction reliability is expected to decline as molecules move beyond the chemistry represented during training. To determine whether ADMET-EvO could recognize this boundary, a post-freeze familiarity score was defined as each test molecule\textquotesingle s maximum fingerprint similarity to the training--validation set. No canonical molecular identity or Bemis--Murcko scaffold overlap was detected between the development and test sets, ensuring that familiarity reflected chemical proximity rather than duplication. Chemical familiarity consistently tracked lower prediction error. After endpoint-wise normalization, the highest-similarity quartile showed lower error in 21 of 22 endpoints, including 12 of 13 classification tasks and all 9 regression tasks (\textbf{Fig. \ref{fig:2}D}). The median error reduction was 15.74\% for classification and 15.39\% for regression. Consistent with this pattern, retaining only the most familiar 25\% of test molecules reduced median normalized error to 85.34\% of the full-test level for classification and 84.14\% for regression. Error returned toward the full-test baseline as coverage expanded, defining an interpretable risk--coverage gradient. Thus, nearest-training-set similarity provides a label-independent applicability signal that allows ADMET-EvO to define prospective, chemically interpretable decision thresholds for prioritization and abstention.

\subsection{ADMET-EvO isolates transferable gains across data, feature and model axes}

\begin{figure}[!htbp]
\centering
\includegraphics[width=\linewidth,height=0.70\textheight,keepaspectratio]{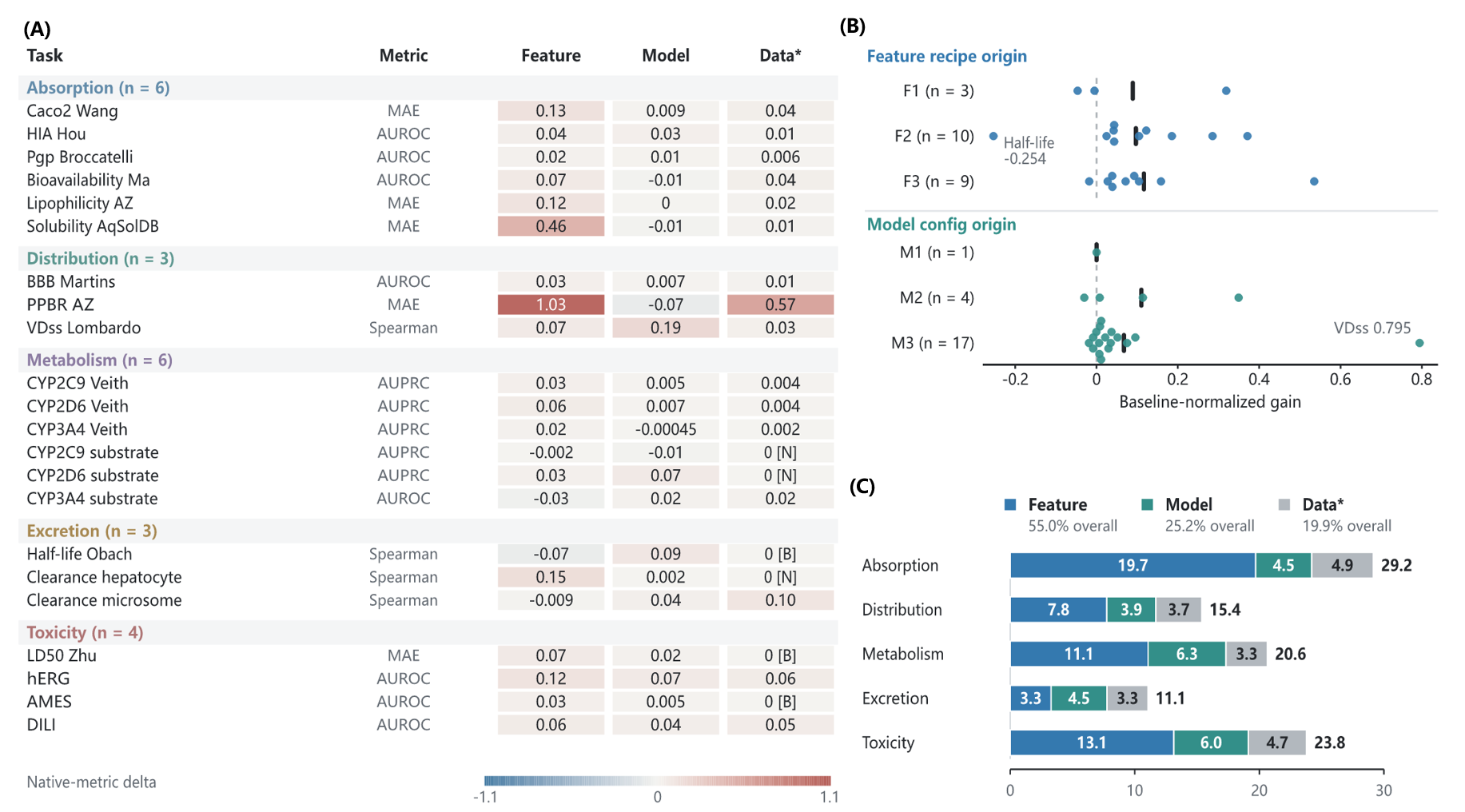}
\caption{\textbf{Three-axis ablation and recipe composition across 22 ADMET tasks.} (A) Oriented task-level changes for Feature, Model and Data*. Positive values indicate improvement. {[}N{]} denotes no eligible external data and {[}B{]} baseline retention (\textbf{Tab.~\ref{tab:S1}}). (B) Baseline-normalized Feature and Model gains grouped by configuration origin. F1--F3 represent successive feature-selection stages, whereas M1--M3 represent model selection, tuning and fusion. Points denote task means across four repeats and black ticks denote group means. (C) Shares of pooled positive gain across ADMET domains after task-wise normalization.}
\label{fig:3}
\end{figure}

The contributions of training data, molecular representation and model selection were examined separately, with the other two components held constant in each ablation. Candidate configurations were compared during development and on an independent validation split, with four different random seeds used at each stage. Each selected configuration was then fixed and assessed with four additional random seeds on the official held-out set. Excluding held-out labels from every selection step enabled a direct test of whether validation gains persisted on unseen compounds. \textbf{Fig. \ref{fig:3}} shows how these gains varied across tasks, which molecular representations and models were selected, and how the three axes contributed across ADMET domains. (\textbf{Tab. \ref{tab:2}})

External training records were selected separately for each endpoint rather than pooled uniformly across tasks. Endpoint-specific ChEMBL 37 cohorts were used for 19 endpoints. BBB, hERG and DILI drew on B3DB, hERG Central and FDA DILIrank2, respectively, and PubChem supplied DILI structures. After compound identity resolution, duplicate removal, label-conflict resolution and leakage filtering, 139,575 candidate records were retained. Sources were matched to the assay context and label definition of each endpoint, maintaining chemical relevance while extending the training space.

During development, the Data axis improved 16 of 22 endpoints, with a mean normalized gain of 3.458\% and a positive median of 1.258\%. The task-level pattern is shown in \textbf{Fig. \ref{fig:3}a}, and the Data axis accounted for 19.9\% of the total positive task-normalized gain with contributions from every ADMET domain (\textbf{Fig. \ref{fig:3}c}). The largest gains were obtained for Clearance microsome (+17.54\%), Caco2 (+13.98\%), PPBR (+7.81\%), hERG (+7.03\%) and DILI (+5.94\%). The five endpoint-specific sets contained 5,082 selected compounds and a total of 4,347 scaffolds when counted separately by endpoint. Median Tanimoto similarity to the nearest compound in the original training set ranged from 0.286 to 0.450. Together, the scaffold counts and similarity distributions showed that the added compounds expanded structural diversity while remaining chemically related to the original training sets. The augmented training sets were fixed after validation and then evaluated on the official held-out set. Of the four positive validation effects, three remained positive on held-out evaluation, while the mean normalized gain across all endpoints remained above zero at 0.603\%. The clearest endpoint-level result was obtained for hERG. Five compounds identified as close analogues of held-out molecules were removed before 1,828 compounds were selected from hERG Central. The selected subsets contained 1,740--1,743 unique scaffolds and had a mean nearest-neighbour Tanimoto similarity of 0.406--0.411 to the original training set. AUROC increased from 0.8551 to 0.8716 on validation and from 0.7319 to 0.8346 on held-out compounds, corresponding to gains of 2.07\% and 14.03\%. The improvement was reproduced across all four held-out evaluation seeds. The larger held-out improvement, together with analogue removal before evaluation, argues against simple duplication as the main explanation. It also shows that endpoint-matched data augmentation retained its validation benefit on held-out compounds.

Alongside the gains from expanded chemical coverage, the largest overall improvements were achieved through endpoint-specific feature selection. With training data and LightGBM held fixed, the Morgan radius-2 baseline was replaced by a feature set selected on validation and fixed before held-out evaluation. The selected feature sets combined complementary chemical descriptors and fingerprints rather than uniformly enlarging the input vector. RDKit2D was included in 14 of 22 configurations and Morgan in 11, while Avalon, ErG and Morgan-count fingerprints each occurred in nine. MACCS keys, the RDKit fingerprint, atom-pair features, topological torsions and Morgan radius 3 were retained more selectively. Multi-view feature sets were selected for 19 endpoints, yielding 19 distinct feature configurations across 22 tasks. This diversity was consistent with endpoint-specific feature requirements.

The endpoint-specific advantage was broadly retained across splits. Positive effects were obtained for 21 of 22 endpoints on the independent validation split, and 18 remained positive on the official held-out test. Thus, 85.7\% of the validation gains retained their direction. Of the 17 feature configurations with reproducible validation gains, 16 retained a held-out effect of at least 0.005, corresponding to 94.1\%. The median normalized gain decreased from 9.55\% to 5.74\% but remained positive. Improvements were obtained for 18 of 22 endpoints on the official test, with a mean normalized gain of 10.38\%. When grouped by selection stage, held-out improvements were obtained for eight of nine complementary-view feature unions, nine of ten composed-view feature sets and one of three single-view feature sets (\textbf{Fig. \ref{fig:3}b}). The Feature axis was the largest contributor in the cross-axis comparison and accounted for 55.0\% of the total positive task-normalized gain (\textbf{Fig. \ref{fig:3}a, c}). Transfer was strongest when feature composition matched endpoint chemistry. For Clearance hepatocyte, Morgan radius-3, ErG, Morgan, the RDKit fingerprint, topological torsion, RDKit2D and Avalon were selected. Spearman correlation was increased by 0.0992 on validation (37.53\%) and by 0.1535 on held-out compounds (53.54\%). For Solubility, MAE was reduced by 0.5711 on validation (41.33\%) and by 0.4644 on held-out evaluation (37.11\%) using Morgan, Avalon, ErG and RDKit2D. RDKit2D alone was selected for Caco2, where the normalized gain increased from 14.54\% on validation to 31.89\% on the held-out set. Larger held-out gains than validation gains were obtained for ten selected feature sets. Transfer of both multi-view feature sets and single descriptors was consistent with endpoint relevance rather than greater feature dimensionality alone.

Additional gains were also obtained through endpoint-specific learner selection. With the original task-specific training data and Morgan representation held fixed, the default LightGBM learner was replaced by an endpoint-specific choice from LightGBM, XGBoost, CatBoost, Random Forest, ExtraTrees, balanced ExtraTrees and RBF-SVM. Seventeen endpoints were assigned ensembles containing two or three model families, while five retained single models. Seventeen configurations were selected during multi-family fusion, four during task-level hyperparameter optimization and one during initial model-family selection (\textbf{Fig. \ref{fig:3}b}). LightGBM, CatBoost, Random Forest, XGBoost and ExtraTrees were retained in 10, 9, 8, 7 and 6 configurations, respectively. Their repeated selection reflected complementarity among learner families rather than dominance by a single model. Model identities, hyperparameters and ensemble weights were fixed before official-test labels were accessed. The held-out comparison therefore evaluated whether a learner selected on validation retained its benefit on unseen compounds.

Model effects were also largely preserved beyond validation. Positive validation effects were obtained for 16 of 22 endpoints, and 13 remained positive on the official held-out set. Thus, 81.3\% of the validation gains retained their direction. Of the 13 model configurations with reproducible validation gains, 11 remained positive on the held-out set, corresponding to 84.6\%. The median normalized gain decreased from 1.64\% to 1.13\% but remained positive. Improvements were recorded for 16 endpoints on the official test, with a mean normalized gain of 7.22\%. These gains spanned the ADMET task panel and accounted for 25.2\% of the total positive task-normalized gain (\textbf{Fig. \ref{fig:3}a, c}).

Strong transfer was obtained across regression and classification endpoints. For VDss, Spearman correlation was increased by 0.1591 on validation and by 0.1932 on held-out compounds using a Random Forest and CatBoost ensemble. These changes corresponded to gains of 74.02\% and 79.46\%. For CYP2D6 substrate prediction, AUPRC improvements of 0.0724 and 0.0706 were retained by a class-weighted Random Forest. For hERG, the AUROC advantage of an XGBoost and ExtraTrees ensemble increased from 0.0164 on validation to 0.0698 on the official test. Because successful models were drawn from different families, no single learner was universally dominant. Across the representation and model axes, 21 of 22 endpoints benefited from at least one form of adaptation. Together, the three axes show that endpoint-relevant training data, molecular representation and model inductive bias provided complementary gains across the ADMET domains.

\begin{table}[htbp]
\centering
\caption{Aggregate validation and held-out gains across the three axes.}\label{tab:2}
\begin{tabular}{@{}lcrrrrrr@{}}
\toprule
& & \multicolumn{2}{c}{Data} & \multicolumn{2}{c}{Feature} & \multicolumn{2}{c}{Model} \\
\cmidrule(lr){3-4}\cmidrule(lr){5-6}\cmidrule(l){7-8}
Suite & N & val & test & val & test & val & test \\
\midrule
TDC & 22 & +0.035 & +0.006 & +0.096 & +0.057 & +0.016 & +0.011 \\
\bottomrule
\end{tabular}
\end{table}

\subsection{ADMET-EvO autonomously expands the task space beyond predefined benchmarks}

\begin{figure}[!htbp]
\centering
\includegraphics[width=\linewidth,height=0.70\textheight,keepaspectratio]{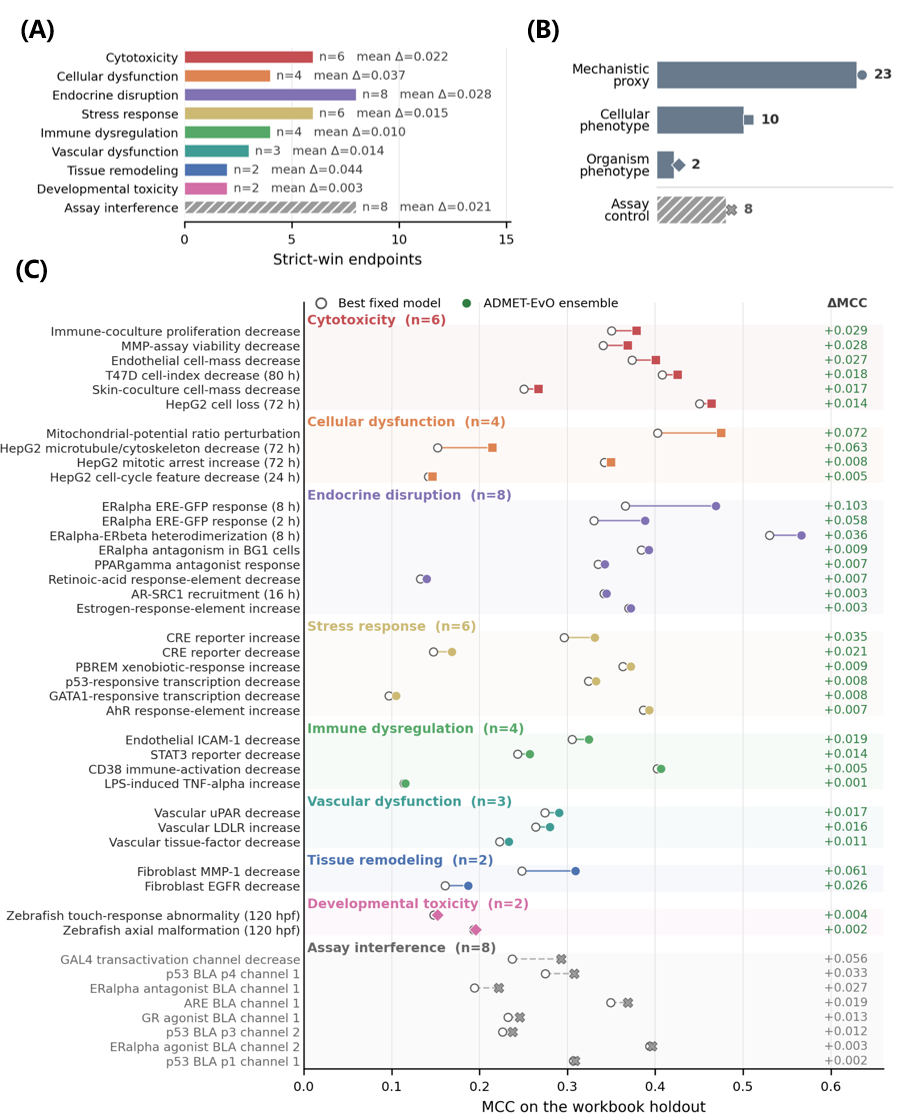}
\caption{\textbf{Autonomous toxicity-task discovery and modelling by ADMET-EvO.} (A) Distribution of 43 endpoint-defined tasks across eight toxicity-relevant classes and a separate assay-control group; annotations indicate task number and mean \ensuremath{\Delta}MCC. (B) Evidence composition comprising 23 mechanistic proxies, 10 cellular phenotypes, 2 organism-level phenotypes and 8 assay controls. (C) Endpoint-wise held-out MCC for ADMET-EvO ensembles (filled symbols) and the best of six fixed single-model pipelines (open circles). Colours denote toxicity class, shapes denote evidence level and grey crosses identify assay controls. Right-hand values indicate \ensuremath{\Delta}MCC = MCC\textsubscript{ADMET-EvO}\ensuremath{-} MCC\textsubscript{best fixed model}. The 43 endpoints represent positive cases with \ensuremath{\Delta}MCC \textgreater{} 0; all values are point estimates. Source data are provided as a Source Data file.}
\label{fig:4}
\end{figure}

The 22-task TDC ADMET benchmark defined ADMET-EvO\textquotesingle s initial knowledge base. ADMET-EvO was subsequently applied to a broader toxicity endpoint space spanning molecular readouts, cellular responses, organism-level phenotypes and assay-validity endpoints. Candidate assay endpoints were not assumed to be model-ready tasks. Instead, each endpoint was parsed into an executable EndpointContract that specified its measured readout, response direction, biological evidence level, assay context, source provenance and compound-level labels. An endpoint was retained as a biologically interpretable task only when its toxicological interpretation and evaluation target could be stated explicitly. Directionally ambiguous or experimentally unoptimized readouts were retained only with an explicit qualification. Single-channel artifact-detection endpoints and internal reporter controls were assigned to the assay-validity category rather than biological toxicity. Applying these criteria enabled ADMET-EvO to autonomously select and formalize 43 endpoint-defined prediction tasks outside its initial knowledge base (\textbf{Fig. \ref{fig:4}a, b}). Task discovery therefore became a structured decision about which endpoints had interpretable biological readouts or necessary assay-validity roles and were sufficiently well defined for computational evaluation.

The 35 biologically interpretable tasks resolved into eight toxicity-relevant classes spanning early pathway perturbation, cellular injury, tissue- and system-level dysfunction, and organism-level phenotypes. Early pathway perturbations comprised endocrine disruption (n = 8) and stress response (n = 6). Endocrine endpoints distinguished receptor assembly (ER\ensuremath{\alpha}--ER\ensuremath{\beta}), co-regulator recruitment (AR--SRC1), antagonism (ER\ensuremath{\alpha} and PPAR\ensuremath{\gamma}), and transcriptional responses (time-resolved ER\ensuremath{\alpha}--ERE activity, an independent ERE reporter and an RAR/RXR-linked direct-repeat-5 response). Together, these mechanistic proxies captured complementary routes to endocrine and metabolic dysregulation rather than repeated measurements of a single estrogenic signal. Stress-response mechanistic proxies captured AhR- and CAR-linked xenobiotic sensing plus p53, bidirectional CRE and GATA signaling, covering enzyme induction, stress adaptation and transcriptional homeostasis. Unoptimized loss-of-signal reporters were retained only with lower interpretive confidence. Cellular evidence comprised cytotoxicity (n = 6) and cellular dysfunction (n = 4). Cytotoxicity was measured as reduced cell number, ATP-based viability, impedance, total protein or proliferation across HepG2, T47D, endothelial, skin and activated endothelial--immune systems. Mitochondrial membrane-potential perturbation, cytoskeletal alteration, mitotic arrest and cell-cycle perturbation identified functional injury that may precede overt cell loss. Tissue- and system-relevant processes comprised immune dysregulation (n = 4), vascular dysfunction (n = 3) and tissue remodeling (n = 2). TNF-\ensuremath{\alpha}, CD38, ICAM-1 and STAT3 mapped inflammatory and immune signaling, whereas uPAR, LDLR and tissue factor mapped vascular migration, lipid homeostasis and coagulation. MMP-1 and EGFR decreases extended this coverage to extracellular-matrix turnover and tissue repair, processes that can mediate or amplify tissue injury. Finally, two developmental-toxicity tasks captured zebrafish axial malformation and abnormal touch response at 120 h post-fertilization, integrating structural and sensorimotor phenotypes. These organism-level readouts were not interpreted as direct evidence of human teratogenicity or neurotoxicity. The remaining eight tasks, seven single-channel \ensuremath{\beta}-lactamase readouts and one GAL4 internal reporter control, were classified separately as assay-validity endpoints. This separation prevented fluorescence or reporter interference from being interpreted as toxicity-relevant pathway activity.

The biological diversity of the selected tasks was matched by diversity in their frozen predictive configurations. ADMET-EvO selected 23 distinct two-pipeline ensemble configurations spanning 13 of the 15 possible pairings among 6 registered base pipelines; the most frequent configuration was used for only 6 tasks. Thirty-three configurations paired 2 pipelines trained on the multiview two-dimensional representation, whereas 10 paired a Morgan-fingerprint pipeline with a multiview two-dimensional pipeline. Seventeen ensembles used equal weights, whereas 26 assigned a 3:1 weight ratio. Representation, base-pipeline pairing and ensemble weighting were therefore selected separately for each endpoint-defined task. On held-out evaluation, every retained ensemble exceeded the retrospective per-endpoint best-of-six fixed single-model comparator (\textbf{Fig. \ref{fig:4}c}). This comparator was stringent because the strongest fixed pipeline was selected separately for each endpoint. Mean Matthews correlation coefficient (MCC) increased from 0.289 to 0.310, corresponding to a mean paired improvement of 0.0218. Among the 35 biologically interpretable tasks, mean MCC reached 0.313; 27 tasks achieved MCC \ensuremath{\geq} 0.20 and 21 achieved MCC \ensuremath{\geq} 0.30. The largest improvements arose from distinct biological processes and distinct ensemble configurations. ER\ensuremath{\alpha}--ERE reporter activity achieved \ensuremath{\Delta}MCC = 0.103 with an ExtraTrees--Random Forest ensemble combining Morgan and multiview two-dimensional representations. Perturbation of mitochondrial membrane potential achieved \ensuremath{\Delta}MCC = 0.072 with a multiview ExtraTrees--XGBoost ensemble. HepG2 microtubule or cytoskeletal alteration achieved \ensuremath{\Delta}MCC = 0.063 with a Morgan ExtraTrees--multiview XGBoost ensemble. Fibroblast MMP-1 protein decrease achieved \ensuremath{\Delta}MCC = 0.061 with an asymmetrically weighted multiview ExtraTrees--Random Forest ensemble. These results showed that the gains were not attributable to one assay family, molecular representation or base pipeline. ADMET-EvO therefore extended beyond model optimization on predefined benchmarks by linking autonomous task selection, explicit endpoint interpretation and endpoint-specific model adaptation within a single workflow.

\section{Methods and Materials}

\subsection{Formalizing the ADMET-EvO research loop}

ADMET-EvO is designed as an endpoint-conditioned autonomous research framework rather than a fixed predictive pipeline. For a given molecular-property endpoint, the system is required to produce not a single validation score but an endpoint-specific predictive recipe together with an auditable research lineage. This lineage records why a recipe was retained, which alternatives were rejected, and what evidentiary strength supports the final claim. Achieving this requires three tightly coupled layers. A task layer fixes what is being predicted. A research layer decides what to try next. An evidence layer decides what the resulting data actually supports.

Different ADMET endpoints encode different biological or physicochemical processes, and modelling choices that are valid for one endpoint need not transfer to another\citep{ref09}. To prevent such assumptions from propagating silently, every endpoint is first compiled into an explicit contract:

\begin{equation}
C =  \langle y, assay, species, label, task, (m, dir), id, prov, split, ext\rangle
\label{eq:1}
\end{equation}

This contract fixes the prediction target \emph{y}, the task type, the primary metric \(m\) and its optimization direction, the molecular-identity convention, data provenance, partition ownership, and the policy governing external evidence. All subsequent reasoning in the system is conditioned on this contract rather than on the raw endpoint description.

Given \(C\), the autonomous research layer maintains a research state that summarizes everything known so far about the endpoint.

\begin{equation}
S_{t} =  \langle C, D_{t}, H_{t}, X, B_{t}\rangle
\label{eq:2}
\end{equation}

Here \(D_{t}\) denotes the current data diagnostics, \(H_{t}\) the history of previously executed interventions, \(X\) the available experts and tools, and \(B_{t}\) the remaining experimental budget. A planning policy \emph{\ensuremath{\pi}} reads \(S_{t}\), formulates a falsifiable hypothesis, and selects one admissible action.

\begin{equation}
a_{t} \in  A_{feature} \cup  A_{model} \cup  A_{data} \cup \{ replicate, stop\}
\label{eq:3}
\end{equation}

This action is drawn from three scientific intervention axes covering molecular feature, predictive model, and training data. Replication and stopping are treated as control actions rather than as additional axes to optimize over. No action reaches execution until it is compiled into a registered, fully specified action.

Execution and judgment are deliberately separated from proposal. A deterministic evaluator \(E\) carries out the selected action and returns a structured evidence record.

\begin{equation}
e_{t} =  E(a_{t}, S_{t}) =  \langle\Delta m_{t}, \sigma_{t}, c_{t}, {status}_{t}, {cfg}_{t}, {prov}_{t}\rangle
\label{eq:4}
\end{equation}

This record comprises the measured effect \(\Delta m_{t}\), its uncertainty or stability \(\sigma_{t}\), computational cost \(c_{t}\) execution status, and full configuration and data provenance. Supported, inconclusive, refuted, and failed outcomes are all retained rather than discarded, so negative results remain part of the endpoint\textquotesingle s evidentiary record. A separate update operator \(U\) then folds this evidence back into the research state.

\begin{equation}
S_{t + 1} =  U(S_{t}, e_{t})
\label{eq:5}
\end{equation}

The update operator selects the next control decision from a fixed set of options. These options are a new intervention, a bounded corrective experiment, retention of the current recipe, rollback to a previously supported configuration, and termination. Together, equations (2) through (5) define a closed research loop.

\begin{equation}
S_t \;\rightarrow\; a_t \;\rightarrow\; e_t \;\rightarrow\; S_{t+1},\qquad \text{driven in turn by }\pi,\ E,\ \text{and }U
\label{eq:6}
\end{equation}

This loop repeats until \emph{U} issues a stop decision. At termination, the system outputs the retained recipe together with the full lineage of supporting and contradicting evidence accumulated along the way. (\textbf{Fig. \ref{fig:S1}a})

The division of labor across equations (3) through (5) is the central design constraint of the framework. The language model instantiates only \emph{\ensuremath{\pi}}, proposing hypotheses and selecting among admissible actions. It never instantiates \(E\) or \(U\). It cannot assign its own evidence verdict, alter the metric \emph{m} or a protected partition, or promote an incomplete experiment to a supported result\citep{ref20,ref27}. Both \(E\) and \(U\) are implemented as deterministic components. The evaluator \(E\) performs data processing, model fitting, metric computation, and protocol checks. The update operator \(U\) performs evidence serialization and state transition. In this architecture, the agent decides what evidence to acquire next. It is the deterministic evaluator and update rule, not the agent, that decide what the acquired evidence supports.

\subsection{Multi-axis closed-loop model and evidence evolution}

Let \emph{T} denote the set of endpoints and let \emph{A} = \{feature, model, data\} denote the three intervention axes, together with two control actions, replicate and stop, that leave the intervention surface unchanged.

\begin{equation}
a =  \langle\tau, axis, c_{ctrl}, c_{trt}, m, seeds, budget, rule\rangle
\label{eq:7}
\end{equation}

A candidate intervention paired a falsifiable hypothesis with an executable action specification identifying the endpoint \emph{\ensuremath{\tau} \ensuremath{\in} T}, the active axis, the control and treatment configurations, the primary metric \(m\), \(a\) fixed set of random seeds, the computational budget, and the promotion or falsification rule. A free-text proposal carried no evidentiary weight until resolved into a registered action \emph{a \ensuremath{\in} A} with a valid configuration and budget, drawn from:

\begin{equation}
a \in  A_{feature}(\tau) \cup  A_{model}(\tau) \cup  A_{data}(\tau) \cup  \{ replicate, stop\}
\label{eq:8}
\end{equation}

Feature actions changed the molecular representation while holding the predictive model and admissible data fixed. Model actions changed the estimator, optimization, regularization, or ensemble configuration while holding representation and data fixed. Data actions changed the empirical evidence available for training through curation, weighting, subset selection, or incorporation of an external source, while preserving the designated model and feature configuration. Replicate actions acquired additional paired seeds to distinguish a stable effect from selection variance, and stop retained the current evidence state once no defensible intervention remained.

For an executed action, the evaluator computed the paired validation effect across the \(K\) random seeds in \(a\),

\begin{equation}
\overline{\Delta}m_{t}(a) =  \frac{1}{K} \sum_{k}^{}{\Delta m_{t}^{(k)}(a)},  \sigma_{t}(a) =  {std}_{k}\lbrack\Delta m_{t}^{(k)}(a)\rbrack
\label{eq:9}
\end{equation}

and assigned one of four verdicts, supported, refuted, inconclusive, or failed, against a fixed decision margin defined by the endpoint contract, before appending the resulting record to the endpoint\textquotesingle s evidence set,

\begin{equation}
H_{t}^{(n + 1)} =  H_{t}^{(n)} \cup  \{ e_{t}(a)\}
\label{eq:10}
\end{equation}

with all four outcomes retained rather than only the best trial.

This verdict was never assigned by the proposing policy. It followed deterministically from \(\overline{\Delta}m_{t}(a)\) and \(\sigma_{t}(a)\) under the metric fixed by the endpoint contract. Holding one axis editable while freezing the other two, and holding the evaluator, baseline, and trial contract fixed, is what lets an observed improvement be traced back to a single intervention axis rather than an unspecified combination of changes.

\subsection{Concrete interventions along the three axes}

A Data action operated on the empirical evidence available to an endpoint before any representation or learner was touched. Each candidate molecule was desalted, neutralized, and reduced to a standardized identity, and any external record matching a protected test, validation, or training molecule under this identity was removed, which caught salt, charge, and tautomer variants that exact-structure matching would miss\citep{ref15,ref28}. An external file was rejected in full when its overlap with an endpoint\textquotesingle s protected molecules exceeded a fixed threshold, since overlap at that scale indicated a shared or derivative source rather than independent evidence. Remaining candidates were further filtered by molecular similarity, with any molecule reaching a fingerprint-based Tanimoto similarity of 0.8 or higher to a protected molecule excluded as a near-analogue\citep{ref29}. (\textbf{Fig. \ref{fig:S1}b}) Sources that passed this firewall were down-weighted relative to native observations through row-level reliability weights and a bounded external-to-native mass cap, and a source that failed independent confirmation was rolled back to the native-data baseline.

A Feature action changed the molecular representation while holding the model and data fixed. The representation portfolio spanned complementary two-dimensional, graph, and three-dimensional views. Fixed two-dimensional representations included Morgan circular fingerprints at multiple radii, count fingerprints, RDKit path fingerprints, atom-pair and topological-torsion fingerprints, MACCS keys, Avalon, ErG, and normalized physicochemical descriptors\citep{ref30,ref31,ref32,ref33,ref34}. Graph encodings and pretrained three-dimensional molecular encoders provided structural and geometric views absent from the fixed fingerprints\citep{ref35,ref36}. Multi-view configurations concatenated a prespecified combination of these blocks after block-specific preprocessing, so that feature interventions ranged from a single fixed encoding to a composed multi-view representation.

A Model action changed the predictive model while holding representation and data fixed. The model portfolio covered tree-based ensembles, including ExtraTrees, random forests, and imbalance-aware variants, gradient-boosted models including CatBoost, XGBoost, and LightGBM, an RBF-support-vector model with a training-size cap, graph neural networks, and pretrained three-dimensional molecular experts, which were activated only when their dependency and compute gates were satisfied\citep{ref35,ref36,ref37,ref38,ref39,ref40,ref41,ref42}. Candidate ensembles combined molecule-level predictions from compatible members, and weighted ensembles were selected per endpoint rather than globally, so the final predictor for one endpoint could differ structurally from another.

Because each axis was varied independently against the same baseline, an observed effect \(\overline{\Delta}m_{t}\) could be attributed to the data available, the representation used, or the model selected, rather than to an unspecified combination of the three.

\subsection{Benchmark tasks and predictive metrics}

The benchmark panel comprised 22 TDC ADMET endpoints, 13 classification tasks and 9 regression tasks, spanning absorption, distribution, metabolism, excretion, and toxicity\citep{ref09}. Each endpoint retained the primary metric and optimization direction specified by its endpoint contract. AUROC was used for most classification tasks, whereas AUPRC was used when pronounced class imbalance made precision--recall performance more informative\citep{ref43}. Regression tasks were evaluated using MAE or Spearman correlation. Spearman correlation was used when relative ordering was more informative than absolute error. These native metrics were retained for endpoint-level reporting.

Because the four metrics differ in scale and direction, cross-task performance was summarized only after normalization within each endpoint. In the equations below, m indexes models, t indexes tasks, x is the reported performance value and b is the best value observed for the same task. For AUROC, AUPRC and Spearman correlation, higher values indicate better performance and the normalized task score was calculated as

\begin{equation}
q_{m,t} =  \frac{x_{m,t}}{b_{t}}
\label{eq:11}
\end{equation}

For MAE, lower values indicate better performance and the ratio was reversed.

\begin{equation}
q_{m,t} =  \frac{b_{t}}{x_{m,t}}
\label{eq:12}
\end{equation}

The overall task-normalized score was then calculated as

\begin{equation}
S_{m} =  \frac{100}{22}\sum_{t = 1}^{22}q_{m,t}
\label{eq:13}
\end{equation}

All 22 endpoints received equal weight. A score of 100 indicates that a model matched the best observed result on every endpoint. The calculation included eight models with results available for all 22 endpoints, and missing values were not imputed. All values entering the aggregation were positive. Five-run means were taken from the corresponding publications and project releases or obtained by direct source-code reproduction. Task-level Top-5 counts followed the native metric direction, and all models tied at fifth place were counted. This aggregate score was used for cross-task comparison, whereas official TDC evaluation remained endpoint-specific.

Endpoint-level performance was compared with published TDC benchmark results under matched task identity and metric direction\citep{ref09}. A separate metric convention was used for tasks generated by ADMET-EvO through autonomous biological interpretation. Because these tasks spanned heterogeneous biological levels, they were evaluated consistently using the Matthews correlation coefficient\citep{ref44}.

The same endpoint-specific metrics were used to quantify data, feature and model interventions against common baselines. Score-only and random selection served as routing controls under matched action budgets. Calibration and conformal coverage were assessed separately from predictive performance and were not included in the task-normalized score\citep{ref45,ref46}.

\subsection{Statistical inference and claim states}

A single evidence record shows whether one intervention helped on one endpoint. A conclusion aggregates many such records, and how that aggregation is done determines how far the result can be trusted. Three rules governed aggregation. The endpoint was the unit of analysis, with seeds treated as repeated measurements rather than independent trials, and a comparison counted only once all prespecified seeds had completed. Metrics on different scales were never averaged directly, only through direction-oriented effects, normalized improvements, or within-task ranks. Configuration selection used only development-stage evidence, and certification results were never used afterward to reselect a configuration. Results were kept in separate tiers, development, confirmation, and certification, since each answers a different question, and a value from one tier was never reported as belonging to another. Failed runs, rejected sources, and inconclusive comparisons stayed in the denominator rather than being dropped. Every reported value carried a claim state, certified, independently confirmed, development-only, exploratory or best-observed, or pending external readout, so the strength of a claim was inferred from its supporting evidence rather than asserted by the agent that produced it.

\section{Conclusions}

ADMET-EvO establishes an evidence-gated foundation for sustained, self-evolving ADMET research. Rather than treating each endpoint as an isolated optimization problem, it maintains a persistent research state in which endpoint definitions, data diagnostics, hypotheses, interventions and evidence verdicts are carried forward across successive cycles. Supported, rejected, inconclusive and failed outcomes are retained to inform subsequent priorities, allowing the system to evolve both its research strategy and its predictive models. Across 22 TDC endpoints, ADMET-EvO achieved the top task-normalized score of 96.77, demonstrating consistently strong performance across heterogeneous prediction settings. Evidence-guided selection reduced cumulative fitting time by 72.2\% under the predefined non-inferiority criterion, while adaptations along the data, feature and model axes produced gains that were largely retained on held-out compounds.

Beyond predefined benchmarks, ADMET-EvO formalized 43 toxicity-related endpoint tasks and connected explicit biological interpretation with endpoint-specific model construction. These results show that the current ADMET-EvO base already possesses strong capabilities for acquiring and evaluating evidence, constructing predictive models and converting accumulated outcomes into decisions about what should be investigated next. Its contribution therefore extends beyond automated model selection: ADMET-EvO organizes individual experiments into a cumulative research programme in which each cycle inherits and refines the knowledge generated by previous cycles.

The present framework thus already realizes continuous evolution at the levels of evidence, task definition, research strategy and model configuration. Further development will extend the temporal horizon, evidence coverage and operational autonomy of this process, allowing ADMET-EvO to continue revising its knowledge and modelling strategies as new data and scientific questions emerge. This persistent, evidence-driven mode of operation positions ADMET-EvO as a foundation for long-horizon autonomous ADMET research rather than a one-off automated modelling workflow.

\FloatBarrier
\nocite{ref01,ref02,ref03,ref04,ref05,ref06,ref07,ref08,ref09,ref10,ref11,ref12,ref13,ref14,ref15,ref16,ref17,ref18,ref19,ref20,ref21,ref22,ref23,ref24,ref25,ref26,ref27,ref28,ref29,ref30,ref31,ref32,ref33,ref34,ref35,ref36,ref37,ref38,ref39,ref40,ref41,ref42,ref43,ref44,ref45,ref46}
\bibliographystyle{admet_references}
\bibliography{references}
\clearpage
\setcounter{figure}{0}
\setcounter{table}{0}
\renewcommand{\thefigure}{S\arabic{figure}}
\renewcommand{\thetable}{S\arabic{table}}
\renewcommand{\theHfigure}{supp.\arabic{figure}}
\renewcommand{\theHtable}{supp.\arabic{table}}
\section*{Supplementary Information}
\addcontentsline{toc}{section}{Supplementary Information}
\begin{center}\large ADMET-EvO: sustained self-evolution across heterogeneous ADMET prediction tasks\end{center}
\begin{figure}[htbp]
\centering
\includegraphics[width=\linewidth]{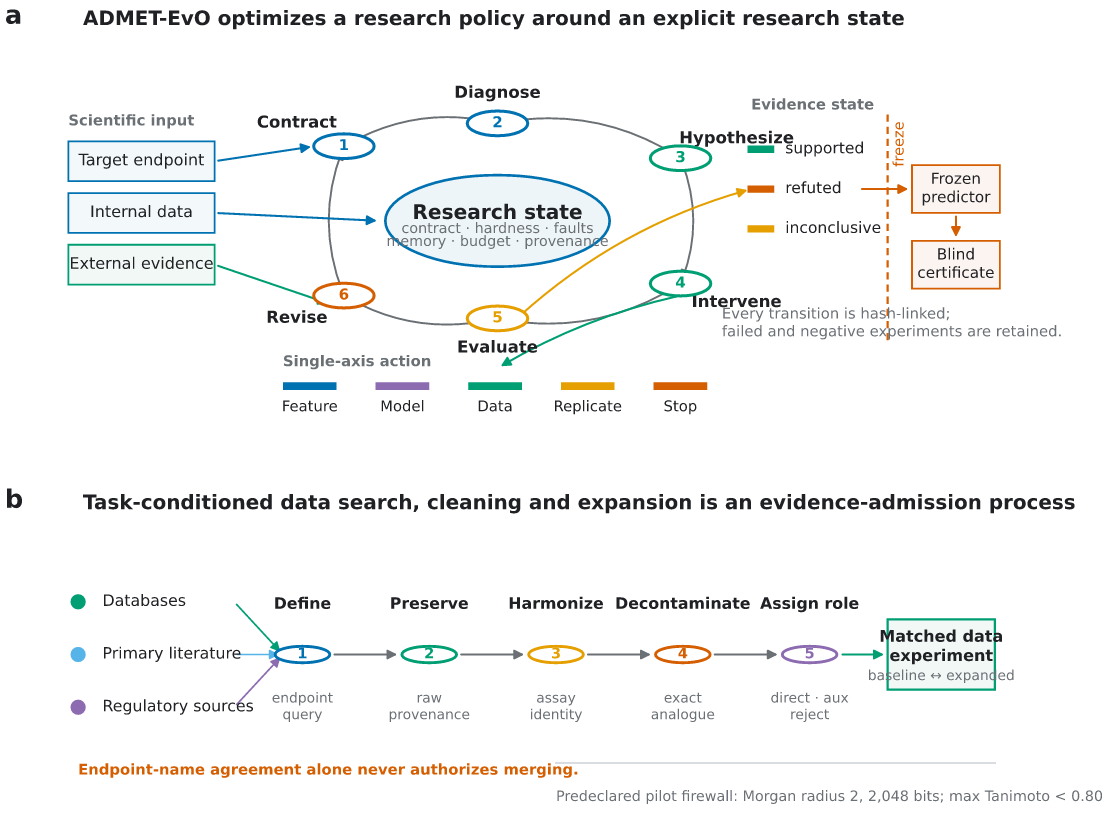}
\caption{Autonomous research architecture of ADMET-EvO. a, Explicit research-state optimization through a contract--diagnose--hypothesize--intervene--evaluate--revise loop, followed by model freezing and blind certification. b, Task-conditioned evidence admission integrating databases, literature and regulatory sources into matched data-expansion experiments.}\label{fig:S1}
\end{figure}
\clearpage
\begingroup
\small
\setlength{\tabcolsep}{4pt}
\renewcommand{\arraystretch}{1.18}
\begin{longtable}{@{}>{\raggedright\arraybackslash}p{\dimexpr 0.29\linewidth-2.32\tabcolsep\relax}>{\raggedright\arraybackslash}p{\dimexpr 0.16\linewidth-1.28\tabcolsep\relax}>{\raggedright\arraybackslash}p{\dimexpr 0.19\linewidth-1.52\tabcolsep\relax}>{\raggedright\arraybackslash}p{\dimexpr 0.18\linewidth-1.44\tabcolsep\relax}>{\raggedright\arraybackslash}p{\dimexpr 0.18\linewidth-1.44\tabcolsep\relax}@{}}
\caption{Raw values underlying the three-axis ablation heatmap}\label{tab:S1}\\
\toprule
Task & Metric & Feature \ensuremath{\Delta} & Model \ensuremath{\Delta} & Data \ensuremath{\Delta} \\
\midrule
\endfirsthead
\multicolumn{5}{l}{\small Table S1 (continued)}\\
\toprule
Task & Metric & Feature \ensuremath{\Delta} & Model \ensuremath{\Delta} & Data \ensuremath{\Delta} \\
\midrule
\endhead
\midrule
\endfoot
\bottomrule
\endlastfoot
Caco2 Wang & MAE & 0.131852 & 0.008940 & 0.039558 \\
HIA Hou & AUROC & 0.040329 & 0.034156 & 0.010000 \\
Pgp Broccatelli & AUROC & 0.022511 & 0.010247 & 0.005593 \\
Bioavailability Ma & AUROC & 0.071749 & -0.012970 & 0.039062 \\
Lipophilicity AZ & MAE & 0.118820 & 0.000000 & 0.016310 \\
Solubility AqSolDB & MAE & 0.464365 & -0.010787 & 0.012406 \\
BBB Martins & AUROC & 0.034152 & 0.006782 & 0.013871 \\
PPBR AZ & MAE & 1.032063 & -0.071893 & 0.570514 \\
VDss Lombardo & Spearman & 0.069827 & 0.193243 & 0.031231 \\
CYP2C9 Veith & AUPRC & 0.032602 & 0.005198 & 0.003649 \\
CYP2D6 Veith & AUPRC & 0.060008 & 0.007398 & 0.003873 \\
CYP3A4 Veith & AUPRC & 0.023548 & -0.000445 & 0.001838 \\
CYP2C9 substrate & AUPRC & -0.002295 & -0.011720 & NA \\
CYP2D6 substrate & AUPRC & 0.026019 & 0.070618 & NA \\
CYP3A4 substrate & AUROC & -0.030289 & 0.018818 & 0.022894 \\
Half-life Obach & Spearman & -0.065704 & 0.090341 & 0.000000 \\
Clearance hepatocyte & Spearman & 0.153525 & 0.002001 & NA \\
Clearance microsome & Spearman & -0.008821 & 0.036476 & 0.103687 \\
LD50 Zhu & MAE & 0.069553 & 0.023190 & 0.000000 \\
hERG & AUROC & 0.116053 & 0.069809 & 0.060470 \\
AMES & AUROC & 0.032422 & 0.005246 & NA \\
DILI & AUROC & 0.060761 & 0.043913 & 0.054674 \\
\end{longtable}
\endgroup
*Values are native-metric oriented differences: \ensuremath{\Delta} = candidate \ensuremath{-} baseline for AUROC, AUPRC and Spearman, and \ensuremath{\Delta} = baseline \ensuremath{-} candidate for MAE; positive values therefore indicate improvement.

\clearpage
\begin{landscape}
\begingroup
\fontsize{8}{10}\selectfont
\setlength{\tabcolsep}{4pt}
\renewcommand{\arraystretch}{1.18}
\begin{longtable}{@{}>{\raggedright\arraybackslash}p{\dimexpr 0.095\linewidth-1.52\tabcolsep\relax}>{\raggedright\arraybackslash}p{\dimexpr 0.12\linewidth-1.92\tabcolsep\relax}>{\raggedright\arraybackslash}p{\dimexpr 0.105\linewidth-1.68\tabcolsep\relax}>{\raggedright\arraybackslash}p{\dimexpr 0.075\linewidth-1.2\tabcolsep\relax}>{\raggedright\arraybackslash}p{\dimexpr 0.08\linewidth-1.28\tabcolsep\relax}>{\raggedright\arraybackslash}p{\dimexpr 0.25\linewidth-4.0\tabcolsep\relax}>{\raggedright\arraybackslash}p{\dimexpr 0.09\linewidth-1.44\tabcolsep\relax}>{\raggedright\arraybackslash}p{\dimexpr 0.085\linewidth-1.36\tabcolsep\relax}>{\raggedright\arraybackslash}p{\dimexpr 0.1\linewidth-1.6\tabcolsep\relax}@{}}
\caption{Final performance and endpoint-specific recipes on the held-out test set.}\label{tab:S2}\\
\toprule
ADMET domain & Endpoint & Task type & Metric & Performance & Model & Feature & Predictor form & Data \\
\midrule
\endfirsthead
\multicolumn{9}{l}{\small Table S2 (continued)}\\
\toprule
ADMET domain & Endpoint & Task type & Metric & Performance & Model & Feature & Predictor form & Data \\
\midrule
\endhead
\midrule
\endfoot
\bottomrule
\endlastfoot
Absorption & Caco-2 Wang & Regression & MAE \ensuremath{\downarrow} & 0.24 & LightGBM & F1 & Single & TDC + external \\
Absorption & HIA Hou & Classification & AUROC \ensuremath{\uparrow} & 100.00\% & 0.5 tuned ExtraTrees + 0.5 baseline ExtraTrees, class-weighted & F1 + F1 & Ensemble & TDC + external \\
Absorption & P-gp Broccatelli & Classification & AUROC \ensuremath{\uparrow} & 91.76\% & Random Forest & F1 & Single & TDC + external \\
Absorption & Bioavailability Ma & Classification & AUROC \ensuremath{\uparrow} & 77.01\% & Fixed CatBoost on MapLight+GNN & F5 & Single & TDC + external \\
Absorption & Lipophilicity AstraZeneca & Regression & MAE \ensuremath{\downarrow} & 0.38 & Fine-tuned Uni-Mol2 84M & U & Single & TDC + external \\
Absorption & Solubility AqSolDB & Regression & MAE \ensuremath{\downarrow} & 0.73 & 0.75 LightGBM(F1) + 0.25 LightGBM(F4) & F1 + F4 & Ensemble & TDC + external \\
Distribution & BBB Martins & Classification & AUROC \ensuremath{\uparrow} & 93.49\% & 0.5 class-weighted RF(F2) + 0.5 class-weighted RF(F1) & F2 + F1 & Ensemble & TDC + external \\
Distribution & PPBR AZ & Regression & MAE \ensuremath{\downarrow} & 6.73 & 0.5 LightGBM(F3) + 0.5 tuned LightGBM(F1) & F3 + F1 & Ensemble & TDC + external \\
Distribution & VDss Lombardo & Regression & Spearman\textquotesingle s \ensuremath{\rho} & 0.70 & Tuned ExtraTrees & F1 & Single & TDC + external \\
Metabolism & CYP2C9 Veith & Classification & AUPRC \ensuremath{\uparrow} & 80.26\% & 0.5 LightGBM + 0.5 tuned XGBoost & F1 + F1 & Ensemble & TDC + external \\
Metabolism & CYP2D6 Veith & Classification & AUPRC \ensuremath{\uparrow} & 72.49\% & 0.5 tuned LightGBM + 0.5 baseline LightGBM, class-weighted & F1 + F1 & Ensemble & TDC + external \\
Metabolism & CYP3A4 Veith & Classification & AUPRC \ensuremath{\uparrow} & 88.68\% & 0.75 tuned XGBoost + 0.25 baseline XGBoost & F1 + F1 & Ensemble & TDC + external \\
\pagebreak[4]
Metabolism & CYP2C9 substrate & Classification & AUPRC \ensuremath{\uparrow} & 43.40\% & Fixed CatBoost on MapLight+GNN & F5 & Single & TDC \\
Metabolism & CYP2D6 substrate & Classification & AUPRC \ensuremath{\uparrow} & 72.10\% & 0.75 tuned class-weighted LightGBM-A + 0.25 tuned class-weighted LightGBM-B & F1 + F1 & Ensemble & TDC \\
Metabolism & CYP3A4 substrate & Classification & AUROC \ensuremath{\uparrow} & 65.59\% & 0.25 RBF-SVM(F2) + 0.75 RBF-SVM(F3) & F2 + F3 & Ensemble & TDC + external \\
Excretion & Half-Life Obach & Regression & Spearman\textquotesingle s \ensuremath{\rho} & 0.59 & Equal-weight rank fusion of fine-tuned Uni-Mol2, OOF top-3 ensemble and rank-target XGBoost & U + dynamic top-3 + F1 & Rank fusion & TDC \\
Excretion & Clearance Hepatocyte AZ & Regression & Spearman\textquotesingle s \ensuremath{\rho} & 0.46 & 0.75 XGBoost + 0.25 LightGBM & F1 + F1 & Ensemble & TDC \\
Excretion & Clearance Microsome AZ & Regression & Spearman\textquotesingle s \ensuremath{\rho} & 0.69 & 0.5 XGBoost(F3) + 0.5 tuned XGBoost(F1) & F3 + F1 & Ensemble & TDC + external \\
Toxicity & LD50 Zhu & Regression & MAE \ensuremath{\downarrow} & 0.59 & 0.75 baseline RF + 0.25 tuned RF & F1 + F1 & Ensemble & TDC \\
Toxicity & hERG & Classification & AUROC \ensuremath{\uparrow} & 92.05\% & 0.5 LightGBM(F1) + 0.5 fine-tuned Uni-Mol2, raw fusion & F1 + U & Raw fusion & TDC + external \\
Toxicity & AMES & Classification & AUROC \ensuremath{\uparrow} & 87.40\% & 0.75 ExtraTrees + 0.25 XGBoost & F1 + F1 & Ensemble & TDC \\
Toxicity & DILI & Classification & AUROC \ensuremath{\uparrow} & 97.47\% & Equal-weight raw fusion of LightGBM, ExtraTrees and fine-tuned Uni-Mol2 & F1 + F1 + U & Raw fusion & TDC + external \\
\end{longtable}
\endgroup
Feature codes: F1 (Rich2D), 2,048-bit Morgan radius-2, 1,024-bit Avalon, ErG and RDKit2D. F2 (Path/Torsion), Morgan radius-2, RDKit path fingerprint, topological torsion and RDKit2D. F3 (Count/Topo), Morgan-count radius-2, AtomPair, MACCS and RDKit2D. F4 (Radius-3), Morgan radius-3, Avalon, ErG and RDKit2D. F5 (MapLight+GNN), 1,024-bit Morgan-count radius-2, 1,024-bit Avalon-count, ErG, selected RDKit2D descriptors and a 300-dimensional GIN supervised-masking embedding. U denotes end-to-end fine-tuned Uni-Mol2 84M without concatenated handcrafted fingerprints.
\end{landscape}

\end{document}